\documentclass[pra, aps, twocolumn, superscriptaddress, nofootinbib, tightenlines, nobibnotes, showpacs, floatfix, 10pt]{revtex4-1}
\usepackage{natbib}
\usepackage{slashed}
\usepackage{graphicx}
\usepackage{subfigure}
\usepackage[usenames, dvipsnames]{color}
\usepackage{graphics}
\usepackage{hyperref}
\usepackage{bm}
\usepackage{amsmath}
\usepackage{xcolor}
\usepackage{amsfonts}
\usepackage{float}

\definecolor{lapislazuli}{rgb}{0.15, 0.38, 0.61}
\definecolor{carmine}{rgb}{0.59, 0.0, 0.09}
\definecolor{carminered}{rgb}{1.0, 0.0, 0.22}

\usepackage{multirow}

\hypersetup{backref,
colorlinks=true,
linkcolor=lapislazuli,
linktoc=page,
citecolor=lapislazuli,
urlcolor=lapislazuli}

\everymath{\displaystyle}

\begin{document}

\title{Hydrodynamic signatures and spectral properties
of the quantum vortex}

\author{Jo\~ao E. H. Braz}

\affiliation{CeFEMA, Instituto Superior T\'ecnico, Universidade de Lisboa, Lisboa,
Portugal}

\author{P. Ribeiro}

\affiliation{CeFEMA, Instituto Superior T\'ecnico, Universidade de Lisboa, Lisboa,
Portugal}

\author{H. Ter\c{c}as}

\affiliation{IPFN, Instituto Superior T\'ecnico, Universidade de Lisboa, Lisboa,
Portugal}

\begin{abstract}
We characterize the low-lying excitations of a quantum vortex in a quasi-two-dimensional Bose-Einstein condensate (BEC) using the standard definition of the density of states (DOS) and a modified version that is sensitive to complementary aspects of the excitation's spectrum. The latter proves to be particularly relevant to studying the polaronic state realized when an impurity is embedded in a quantum vortex. We establish that the impurity becomes sensitive to the transversal fluctuations of the vortex, via its remnant kelvon mode, and to the phase fluctuations of the BEC Nambu-Goldstone mode. The presence of the vortex yields an anomalous excitation spectrum with a finite energy gap and non-linear DOS at low energies. We find that the high sensitivity of the kelvon mode to external potentials provides a channel of quantum-level control over impurities trapped in a vortex. This extra control channel may be of practical use for the proposal of using vortex-trapped impurities as qubit units for quantum information processing.
\end{abstract}
\maketitle

\section{Introduction}

Bose-Einstein condensates (BEC) host a variety of topological defects emerging collectively from the interactions among its constituent atoms. Notable examples are vortex configurations occurring in three-dimensional and in quasi-two-dimensional (quasi-2D) BECs. Their vorticity is quantized and the singularity in the phase of the BEC’s order parameter causes the fluid to deplete with a long-range density profile~\cite{Pitaevskii:1961aa}. Although a single-charge vortex is dynamically stable and topologically protected from dissipation~\cite{PhysRevA.70.043610}, the system is bound up with multiple dynamics. For one, the vortex is intrinsically subject to the transversal excitations of its filament, known as kelvon modes; these can precipitate the vortex off its rotational axis, triggering an outwards precessional motion that eventually drives it to the boundary~\cite{Dalfovo:1996aa,Rokhsar:1997aa,Isoshima:1997aa,Dodd:1997aa,Fetter:1998aa,Isoshima:1999ac,Fedichev:1999aa,svidzinsky_2000,Virtanen:2001ab}. Once the vortex is removed then so is the phase singularity, and the BEC is left to evolve according to its intrinsic dynamics alone: the fluid undergoes quantum diffusion~\cite{Lewenstein:1996aa}, driven by the Nambu-Goldstone (NG) mode inherent to the BEC state. One of our main results is to show how these two modes—a kelvon and the NG mode—have also an exceptional capacity to drive the dynamics of impurities embedded in the quantum fluid.

The physics of a BEC in the presence of impurities (e.g., heterogeneous atomic species) has aroused interest in both theory and experiment over the past decade~\cite{tempere_2009,PhysRevLett.117.055302,yoshida_2018,drescher_2019,mistakidis_2019,ardila_2019,Hryhorchak:2020aa,Khan:2021ui}; of particular interest are cases when the BEC hosts one or multiple vortices, motivated by the possibility of impurities becoming bounded in their core or hopping across the Abrikosov lattice~\cite{Johnson:2016aa,Braz:2020aa,Edmonds:2020aa,Richaud:2020ut}. These systems naturally give rise to problems of quantum many-body physics, specifically, along the branch of polaron physics~\cite{Alexandrov:2010aa,Emin:2012aa}; spectral properties of such many-body systems are ubiquitously extracted via a variety of measures broadly referred to as densities of states—energy-distributions of excited states. We introduce and analyze a class of densities of states, which we call \emph{hydrodynamic}, that naturally arise in this problem and which characterize the capacity of each excitation to interact with a given impurity. Moreover, we analyze the standard densities of states of Bogoliubov excitations to resolve their spectral structure. Remarkably, the latter point to the existence of anomalies in the excitation spectrum at small momenta, which are rooted in the long-range profile of the vortex.

In this work, we provide a detailed investigation of the low-lying spectrum of excitations of a quasi-2D BEC supporting a single-charge, on-axis vortex (vortex-BEC) by means of densities of states (DOS) and local densities states (LDOS) alike. The hydrodynamic LDOS, introduced here, follow from the leading-order coupling of an impurity to the excitations of a BEC. These measures show the exceptional sensitivity of an impurity to the remnant kelvon mode~\cite{Simula:2018aa}, a.k.a. the lowest core-localized state (LCLS)~\cite{Isoshima:1999ac}, and to the Nambu-Goldstone mode of the spontaneous symmetry-breaking of the phase degree-of-freedom of a BEC. We show that the action of a pinning potential, which we include as the stabilization mechanism of the vortex~\cite{Isoshima:1999ac,Isoshima:1999ab}, affords a degree of quantum-level control over an impurity trapped in the core of the vortex by medium of the LCLS.

The paper is organized as follows: in Sec. ~\ref{subsec:Bogoliubov-equations-for}, we review the Bogoliubov formalism applied to a vortex-BEC. We compute the relevant densities of states in Sec.~\ref{subsec:Local-densities-of}, where we introduce the Bogoliubov and hydrodynamic DOS, followed by a description of the physical setup in Sec.~\ref{subsec:Trapping-and-pinning}. Numerical results are shown in Sec.~\ref{sec:Results}, where we highlight the spectral features of the vortex-BEC by comparison with the homogeneous, zero-vorticity case. Conclusion and discussion follow in Sec.~\ref{sec:Discussion-and-Conclusions}.

\section{Formulation}


\subsection{Bogoliubov formalism for a vortex-BEC\label{subsec:Bogoliubov-equations-for}}

We begin by considering the field-quantized Hamiltonian

\begin{align}
\hat{H} & =\int\text{d}^{2}r\,\hat{\Phi}^{\dagger}(\mathbf{r})\left[-\frac{\hbar^{2}}{2M}\nabla^{2}-\mu+V(\mathbf{r})\right]\hat{\Phi}(\mathbf{r})\nonumber \\
 & +\frac{1}{2}g\int\text{d}^{2}r\,\hat{\Phi}^{\dagger}(\mathbf{r})\hat{\Phi}^{\dagger}(\mathbf{r})\hat{\Phi}(\mathbf{r})\hat{\Phi}(\mathbf{r})\,,\label{eq:Form_Ham}
\end{align}
describing the field of a bosonic species of mass $M$ interacting
by an effective contact potential of strength $g>0$. Atomic annihilation
and creation operators $\hat{\Phi}$ and $\hat{\Phi}^{\dagger}$ satisfy
the commutation relation $[\hat{\Phi}(\mathbf{r}),\hat{\Phi}^{\dagger}(\mathbf{r}')]=\delta(\mathbf{r}-\mathbf{r}')$
and $\hat{N}=\int\text{d}^{2}r\,\hat{\Phi}^{\dagger}(\mathbf{r})\hat{\Phi}(\mathbf{r})$
is the number operator, whose expectation value is fixed by the chemical
potential $\mu$. The external potential $V$ is comprised of two
contributions: a highly anisotropic trapping potential $V_{\text{tr}}$,
rendering the system quasi-2D; a small pinning potential $V_{\text{p}}$,
introduced here to energetically stabilize the vortex. (These potentials
are detailed in Sec.~\ref{subsec:Trapping-and-pinning}.)

The Bogoliubov approach suffices in the case of a weakly interacting
BEC~\cite{Dalfovo:1999aa}. We thus decompose the field as
\begin{equation}
\left(\begin{array}{c}
\hat{\Phi}(\mathbf{r})\\
\hat{\Phi}^{\dagger}(\mathbf{r})
\end{array}\right)=\sqrt{n_{0}}\left(\begin{array}{c}
\Phi_{0}(\mathbf{r})\\
\overline{\Phi_{0}}(\mathbf{r})
\end{array}\right)+\left(\begin{array}{c}
\hat{\phi}(\mathbf{r})\\
\hat{\phi}^{\dagger}(\mathbf{r})
\end{array}\right)\,,\label{eq:Form_Bogol-presc}
\end{equation}
where $\Phi_{0}$ is a BEC wave function ($\overline{\Phi_{0}}$ its
complex-conjugate), on top of which $\hat{\phi}(\mathbf{r})$ ($\hat{\phi}(\mathbf{r})^{\dagger}$)
annihilates (creates) quantum excitations. Here, $n_{0}=N_{0}/\mathcal{A}$,
with $N_{0}$ the number of BEC atoms and $\mathcal{A}$ the area
covered by the quasi-2D BEC cloud. The mean-field $\Phi_{0}$ satisfies
the time-independent Gross-Pitaevskii (GP) equation,

\begin{equation}
\left(h-\mu+n_{0}g\left|\Phi_{0}\right|^{2}\right)\Phi_{0}=0\,,\label{eq:Form_GP-eq}
\end{equation}
with $h=-(\hbar^{2}/2M)\nabla^{2}+V$, subject to the boundary condition
consistent with the external potential $V$ and to the normalization
condition 
\begin{equation}
\bigl\langle\hat{N}\bigr\rangle_{0}=n_{0}\int\text{d}^{2}r\,\left|\Phi_{0}(\mathbf{r})\right|^{2}=N_{0}\,,\label{eq:Form_GP-normalization}
\end{equation}
where $\left\langle \text{\dots}\right\rangle _{0}$ represents the
expected value in the many-body state of the BEC, which can be used
to determine $\mu$.\textcolor{blue}{{} }

We choose the natural microscopic units to be the coupling energy
$n_{0}g$ and the coherence length $\xi=\hbar/\sqrt{2Mn_{0}g}$, so
that quantities in the Hamiltonian~\eqref{eq:Form_Ham} become rescaled
as $\mathbf{r}\mapsto\xi\mathbf{r}$, $\hat{\Phi}\mapsto\xi^{-1}\hat{\Phi}$,
$\mu\mapsto\left(n_{0}g\right)\mu$, $h\mapsto\left(n_{0}g\right)h$
and $\hat{H}\mapsto\left(n_{0}\xi^{2}\right)\left(n_{0}g\right)\hat{H}$;
substituting for~\eqref{eq:Form_Bogol-presc} and by virtue of~\eqref{eq:Form_GP-eq},
we have
\begin{equation}
\hat{H}=F_{0}+\frac{1}{n_{0}\xi^{2}}\hat{H}_{\text{B}}+\text{\dots}\label{eq:Form_Ham-expand}
\end{equation}
where $F_{0}=\int\text{d}^{2}r\,\overline{\Phi_{0}}\left(h-\mu+\frac{1}{2}\left|\Phi_{0}\right|^{2}\right)\Phi_{0}$
is the (classical) free energy of the BEC and
\begin{equation}
\hat{H}_{\text{B}}=\frac{1}{2}\int\text{d}^{2}r\,\left(\begin{array}{c}
\hat{\phi}\\
\hat{\phi}^{\dagger}
\end{array}\right)^{\dagger}\sigma_{3}\mathcal{H}_{\text{B}}\left(\begin{array}{c}
\hat{\phi}\\
\hat{\phi}^{\dagger}
\end{array}\right)\label{eq:Form_Ham-BdG}
\end{equation}
with the Bogoliubov operator
\begin{equation}
\mathcal{H}_{\text{B}}=\sigma_{3}\left(\begin{array}{cc}
h-\mu+2\left|\Phi_{0}\right|^{2} & \Phi_{0}^{2}\\
\overline{\Phi_{0}}^{2} & h-\mu+2\left|\Phi_{0}\right|^{2}
\end{array}\right)~,\label{eq:Form_BdG-op}
\end{equation}
where $\sigma_{i}$ $(i=1,2,3)$ denote Pauli
matrices. Higher-order contributions of Eq.~\eqref{eq:Form_Ham-expand}
are neglected within the Bogoliubov approximation. This requires $\bigl\langle\hat{N}\bigr\rangle_{0}\approx\bigl\langle\hat{N}\bigr\rangle$,
which holds provided $k_{B}T\ll\mu$ and $n_{0}\xi^{2}\gg1$, i.e.,
in the limit of both weak thermal and quantum depletions. We then
expand the excitation operators as
\begin{equation}
\left(\begin{array}{c}
\hat{\phi}(\mathbf{r})\\
\hat{\phi}^{\dagger}(\mathbf{r})
\end{array}\right)=\sum_{\lambda=0}^{\infty}\left(X_{\lambda}(\mathbf{r})\hat{b}_{\lambda}+\sigma_{1}\overline{X_{\lambda}}(\mathbf{r})\hat{b}_{\lambda}^{\dagger}\right)\,,\label{eq:Form_fluct-expand}
\end{equation}
in a complete basis of complex vector-valued functions $X_{\lambda}(\mathbf{r})=\left(u_{\lambda}(\mathbf{r}),v_{\lambda}(\mathbf{r})\right)^{T}$
($\overline{X_{\lambda}}$ its complex-conjugate), whose components
$u_{\lambda}$ and $v_{\lambda}$ are the particle and hole components,
respectively. By requiring that each $\hat{b}_{\lambda}$ and $\hat{b}_{\lambda}^{\dagger}$
inherit bosonic commutation relations, the diagonalization of~\eqref{eq:Form_Ham-BdG}
becomes an eigenproblem for the $X_{\lambda}$ as 
\begin{equation}
\mathcal{H}_{\text{B}}X_{\lambda}=\omega_{\lambda}X_{\lambda}\,.\label{eq:Form_BdG-eq}
\end{equation}
Bogoliubov operators $\mathcal{H}_{\text{B}}$ are non-hermitian and, in general,
admit complex eigenvalues~\cite{PhysRevA.70.043610}; the operator we study presently, however, possesses a fully real spectrum~\cite{Isoshima:1997aa} and the bi-orthonormality relations
\begin{equation}
\left\langle X_{\lambda},X_{\lambda'}\right\rangle =\int\text{d}^{2}r\,X_{\lambda}^{\dagger}\sigma_{3}X_{\lambda'}=\delta_{\lambda,\lambda'}\,,\label{eq:Form_BdG-ortho}
\end{equation}
\begin{equation}
\left\langle \sigma_{1}\overline{X_{\lambda}},X_{\lambda'}\right\rangle =\int\text{d}^{2}r\,\left(\sigma_{1}\overline{X_{\lambda}}\right)^{\dagger}\sigma_{3}X_{\lambda'}=0\,,\label{eq:Form_BdG-ph}
\end{equation}
hold for every normalized eigenstate, where we introduce the bilinear
product $\left\langle \cdot,\cdot\right\rangle $; the diagonalized
form of~\eqref{eq:Form_Ham-BdG} thus reads
\begin{equation}
\hat{H}_{\text{B}}=\frac{1}{2}\alpha\hat{P}^{2}+\sum_{\lambda\neq0}\omega_{\lambda}\hat{b}_{\lambda}^{\dagger}\hat{b}_{\lambda}\,,\label{eq:Form_BdG-diag}
\end{equation}
where the first term accounts for the dynamics of the phase degree-of-freedom
of the BEC~\cite{Lewenstein:1996aa}.
We explain the structure and importance of this term by observing
that the variation of Eq.~\eqref{eq:Form_GP-normalization}, i.e.,
$\delta\bigl\langle\hat{N}\bigr\rangle_{0}=\delta N_{0}$, yields
the condition
\begin{equation}
\int\frac{\text{d}^{2}r}{\mathcal{A}}\,\left(\overline{\Phi_{0}}\Theta_{0}+\Phi_{0}\overline{\Theta_{0}}\right)=1\,,\label{eq:Form_BdG-adj-cond}
\end{equation}
where we introduce the adjoint BEC wave function
\begin{equation}
\Theta_{0}(\mathbf{r})=\left(N_{0}\frac{\partial\mu}{\partial N_{0}}\right)\frac{\text{d}\Phi_{0}}{\text{d}\mu}(\mathbf{r})\,.\label{eq:Form_BdG-adj}
\end{equation}
It is because Eq.~\eqref{eq:Form_GP-eq} holds that $\Phi_{0}$ and
$\Theta_{0}$ comprise, respectively, the zero-mode of $\mathcal{H}_{\text{B}}$
(the Nambu-Goldstone mode of the BEC state~\cite{Matsumoto:2002aa})
and its adjoint, as
\begin{equation}
\mathcal{H}_{\text{B}}\left(\begin{array}{c}
\Phi_{0}\\
-\overline{\Phi_{0}}
\end{array}\right)=0\,,\quad\mathcal{H}_{\text{B}}\left(\begin{array}{c}
\Theta_{0}\\
\overline{\Theta_{0}}
\end{array}\right)=\alpha\left(\begin{array}{c}
\Phi_{0}\\
-\overline{\Phi_{0}}
\end{array}\right)\,.\label{eq:Form_BdG-zero-mode}
\end{equation}
The second of these equations is obtained by taking the derivative
$\text{d}/\text{d}\mu$ of the first and then multiplying it by $\alpha$,
where we identify $\alpha\equiv N_{0}(\partial\mu / \partial N_{0})$~\cite{Lewenstein:1996aa}.
These solutions make up the $\lambda=0$ mode in \eqref{eq:Form_fluct-expand} as $X_{0}=\left(\Theta_{0}+\Phi_{0},\overline{\Theta_{0}}-\overline{\Phi_{0}}\right)^{T}/\sqrt{2\mathcal{A}}$,
which is normalized, per Eq.~\eqref{eq:Form_BdG-adj-cond}, as $\left\langle X_{0},X_{0}\right\rangle =1$,
thus determining $\alpha$. The associated operators $\hat{b}_{0}$
and $\hat{b}_{0}^{\dagger}$
comprise the Hermitian operator $\hat{P}=\frac{1}{\sqrt{2}}(\hat{b}_{0}+\hat{b}_{0}^{\dagger})$~\cite{Blaizot:1986aa}.
For conciseness, we refer to the mode $X_{0}$ as the Nambu-Goldstone
(NG) mode henceforth. The basis~\eqref{eq:Form_fluct-expand} thus
becomes complete. (The zero-mode pertaining to the position of the
vortex core, though related to the present discussion, is present
only when the vortex is off-axis~\cite{Wright:2009aa}.)

Making the potential $V$ be isotropic in the plane and assuming a
vortex to be on the axis of a disk-shaped quasi-2D BEC, a vortex solution
of the GP Eq.~\eqref{eq:Form_GP-eq} can be written in polar coordinates
as
\begin{equation}
\Phi_{\nu}(r,\varphi)=e^{\text{i}\nu\varphi}\phi_{\nu}(r)\,,\label{eq:Vortex_single-vortex-sol}
\end{equation}
where $\nu$ is an integer, the quantized vorticity, and
$\phi_{\nu}$ satisfies a reduced form of Eq.~\eqref{eq:Form_GP-eq}
in the radial coordinate~\cite{Manton:2004aa}. Inserting~\eqref{eq:Vortex_single-vortex-sol}
in Eq.~\eqref{eq:Form_BdG-op}, it becomes apparent that solutions
of Eq.~\eqref{eq:Form_BdG-eq} can be separated, in polar coordinates,
as
\begin{equation}
X_{\lambda}(r,\varphi)=e^{\text{i}\varphi\left(m+\nu\sigma_{3}\right)}X_{m,n}(r)\,,\label{eq:Vortex_Bogol-amplitudes}
\end{equation}
where $m$ is an integer, the angular momentum of the excitation,
and $n$, the number of nodes in the radial coordinate, is the quantum
number associated to the non-zero eigenstates of the reduced Bogoliubov
operator
\begin{equation}
\sigma_{3}\mathcal{H}_{m}=\left(\begin{array}{cc}
h-\mu+2\phi_{\nu}^{2} & \phi_{\nu}^{2}\\
\phi_{\nu}^{2} & h-\mu+2\phi_{\nu}^{2}
\end{array}\right)+\frac{1}{r^{2}}\left(m+\nu\sigma_{3}\right)^{2}\,,\label{eq:Vortex_BdG-reduced}
\end{equation}
acting on $X_{m,n}(r)=\left(\alpha_{m,n}(\mathbf{r}),\beta_{m,n}(\mathbf{r})\right)^{T}$.
The label $\lambda\neq0$ is, thus, explicitly identified with the
pair of quantum numbers $(m,n)$ of solutions of the eigenproblem
\begin{equation}
\mathcal{H}_{m}X_{m,n}=\omega_{m,n}X_{m,n}\,.\label{eq:Vortex_BdG-eq-reduced}
\end{equation}
Note that, by Eqs.~\eqref{eq:Form_BdG-adj} and~\eqref{eq:Form_BdG-zero-mode},
the NG mode has angular momentum $m=0$.

\subsection{Densities of states\label{subsec:Local-densities-of}}

The local density of states (LDOS) of Bogoliubov excitations is typically defined in terms of the imaginary part of the
associated Green's function~\cite{Mahan:2013aa}. In particular,
we can consider an angular momentum-resolved LDOS (AM-LDOS), given
by
\begin{equation}
D_{m}^{(\text{B})}(\mathbf{r};\omega)=\sum_{n}\varrho_{m,n}^{(\text{B})}(\mathbf{r})\delta(\omega-\omega_{m,n})\,,\label{eq:LDOS_Bogol-ldos}
\end{equation}
with the spatial density $\varrho_{m,n}^{(\text{B})}$ given by
\begin{equation}
\varrho_{m,n}^{(\text{B})}(\mathbf{r})=\left|u_{m,n}(\mathbf{r})\right|^{2}-\left|v_{m,n}(\mathbf{r})\right|^{2}\,.\label{eq:LDOS_Bogol-spat-dens}
\end{equation}
However, a test particle of a different species (i.e., an atomic impurity) that interacts with the species comprising the BEC can be
found to be sensitive to AM-LDOS of a distinct form:
\begin{equation}
D_{m}^{(\text{H})}(\mathbf{r};\omega)=\sum_{n}\varrho_{m,n}^{(\text{H})}(\mathbf{r})\delta(\omega-\omega_{m,n})\,,\label{eq:LDOS_hydro-ldos}
\end{equation}
with the spatial density $\varrho_{m,n}^{(\text{H})}$ given by
\begin{equation}
\varrho_{m,n}^{(\text{H})}(\mathbf{r})=\left|\overline{\Phi}_{0}(\mathbf{r})u_{m,n}(\mathbf{r})+\Phi_{0}(\mathbf{r})v_{m,n}(\mathbf{r})\right|^{2}\,,\label{eq:LDOS_hydro-spat-dens}
\end{equation}
which is due, physically, to the interactions with the bosonic system
being exclusively density interactions. The derivation of this AM-LDOS
is illustrated in Appendix~\eqref{sec:Fermi-golden-rule}. We shall
refer to the quantities in, and derived from, Eqs.~\eqref{eq:LDOS_Bogol-ldos}
and~\eqref{eq:LDOS_hydro-ldos} as the Bogoliubov and the hydrodynamic
densities of states, respectively. We coin the latter
\emph{hydrodynamic} densities due to the $\varrho_{m,n}^{(\text{H})}$
being identical to the density degree of freedom used in the hydrodynamic
formalism for the excitations of a BEC~\cite{Dalfovo:1999aa}. 

Associated to the LDOS is the density of states of states (DOS), which provides a measure for counting excitations of a many-body system~\cite{Dalfovo:1999aa}. The angular momentum-resolved DOS (AM-DOS) are obtained by integrating in space each of the AM-LDOS, yielding
\begin{equation}
D_{m}^{(O)}(\omega)=\sum_{n}c_{m,n}^{(O)}\delta(\omega-\omega_{m,n})\,,\label{eq:LDOS_am-dos}
\end{equation}
where $O=\text{B},\text{H}$, with weights $c_{m,n}^{(\text{B})}=1$
for the Bogoliubov densities of states, by virtue of Eq.~\eqref{eq:Form_BdG-ortho},
and
\begin{equation}
c_{m,n}^{(\text{H})}=\int\text{d}^{2}r\,\varrho_{m,n}^{(\text{H})}(\mathbf{r})\label{eq:LDOS_Hydro-wght}
\end{equation}
for the hydrodynamic density of states. The (total) LDOS and DOS are
recovered upon summing over all $m$, i.e.,
\begin{equation}
D^{(O)}(\omega)=\sum_{m}D_{m}^{(O)}(\omega)\label{eq:LDOS_Bogol-dos}
\end{equation}
for the DOS; as in Eq.~\eqref{eq:LDOS_am-dos}, this is for $O=\text{B},\text{H}$.

For comparison with numerical results, we consider a large, homogeneous
(i.e., vorticity zero and no boundary effects) quasi-2D BEC, in which
case excitations have the Bogoliubov spectrum $\omega_B(\mathbf{k})=\sqrt{k^{2}\left(2\mu+k^{2}\right)}$,
with $\mathbf{k}$ the mode's wave-vector, or momentum, and $k=\left\Vert \mathbf{k}\right\Vert $.
The Bogoliubov and the hydrodynamic DOS of excitations of the homogeneous
quasi-2D BEC can then be explicitly computed in the continuum approximation
of momentum space, i.e., approximating the sum over modes by an integral
over $\mathbf{k}$, yielding
\begin{equation}
D_{B}^{(\text{B})}(\omega)=\frac{A}{4\pi}\frac{\omega}{\sqrt{\mu^{2}+\omega^{2}}}=\frac{A}{4\pi}\frac{\omega}{\mu}-\mathcal{O}(\omega^{3})\,,\label{eq:LDOS_continuum-Bogol-dos}
\end{equation}
\begin{equation}
D_{B}^{(\text{H})}(\omega)=\frac{\mu A}{4\pi}\left(1-\frac{\mu}{\sqrt{\mu^{2}+\omega^{2}}}\right)=\frac{A}{8\pi}\frac{\omega^{2}}{\mu}-\mathcal{O}(\omega^{4})\,,\label{eq:LDOS_continuum-Hydro-dos}
\end{equation}
with $A=\mathcal{A}/\xi^{2}$ the area covered by the BEC in natural
units.

\subsection{Vortex nucleation and external potentials\label{subsec:Trapping-and-pinning}}

We envision the on-axis vortex to be nucleated by phase imprinting
via Laguerre-Gauss beams~\cite{PhysRevLett.97.170406}, that is,
without imposing any laboratory-frame rotation on the fluid.

We consider the anisotropy of the trapping potential $V_{\text{tr}}$
to be produced by a tight harmonic potential in the $z$-direction
of energy $\hbar\omega_{z}\gg\mu$, yielding the effective interaction
strength $g=2\sqrt{2\pi}\hbar^{2}a/(Ml_{z})$, where $a>0$ is the
$s$-wave scattering length and $l_{z}=\sqrt{\hbar/(M\omega_{z})}$
is the characteristic length of the harmonic potential.

The in-plane radial trap is a box potential of radius $\mathcal{R}\gg\xi$,
making Eq.~\eqref{eq:Form_GP-eq} subject to the boundary condition
$\Phi_{0}\left(\mathcal{R},\varphi\right)=0$. In natural units, the
radius $\mathcal{R}=R\xi$ is given by $R=\sqrt{N/(\pi n_{0}\xi^{2})}$,
where $n_{0}\xi^{2}=l_{z}/(4\sqrt{2\pi}a)$ is independent of the
in-plane radial size ($R$ is related to a dimensionless coupling
strength introduced in Refs.~\cite{Rokhsar:1997aa} and~\cite{Butts:1999aa}).
For the pinning potential, we consider $V_{\text{p}}(r)=\epsilon\exp\left(-r^{2}/w^{2}\right)$,
i.e., a Gaussian beam with maximum optical potential $\epsilon$ and
waist length $w$~\cite{Isoshima:1999ac,Ryu:2007aa}; we shall refer
to pinning configurations in terms of the ordered pair $(\epsilon,w)$. 

For the numerical calculations, we consider experiments with $^{7}\text{Li}$
BECs~\cite{Gross:2008aa} wherein positive values of $a$ of tens
of nanometers are accessible via Feshbach resonances, while typical
trapping frequencies of hundreds of kiloHertz yield $l_{z}$ of hundreds
of nanometers~\cite{zhang_2008}, resulting $n_{0}\xi^{2}\gg1$
and, thus, guaranteeing a regime of negligible quantum depletion consistent
with the Bogoliubov approximation. Moreover, in these conditions,
a BEC of up to $N\approx10^{5}$ atoms may yield up to $R\approx200$,
while experiments with box potentials up to $\mathcal{R}\approx70\mu\text{m}$~\cite{gotlibovych_2014}
yield, in laboratory units, $n_{0}g/\hbar\approx30\text{kHz}$, or
$n_{0}g/k_{\text{B}}\approx200\text{nK}$, and $\xi\approx300\text{nm}$.

\section{Results\label{sec:Results}}

We give an account of a numerical analysis of the quantities presented
in Sec.~\ref{subsec:Local-densities-of}. We begin by briefly explaining
the numerical methods used and the motivation for the inclusion of
a pinning potential. Then, we present and discuss results for the
Bogoliubov and the hydrodynamic DOS, Eq.~\eqref{eq:LDOS_Bogol-dos}
for $O=\text{B},\text{H}$, where we will encounter details that motivate
an investigation of the low-lying states of definite angular momentum. Using
the AM-DOS and the AM-LDOS, we make the physical origin of the hydrodynamic
signatures clear. We follow up with a detailed account of the dependence
of low-energy modes on the pinning potential showing, in particular,
the sensitivity of the LCLS and of its hydrodynamic weight to this
perturbation. Finally, we provide a scaling analysis by which we identify anomalies in the low-lying excitation spectrum.

\subsection{Numerical vortex solutions}

\begin{figure}
\begin{centering}
\includegraphics{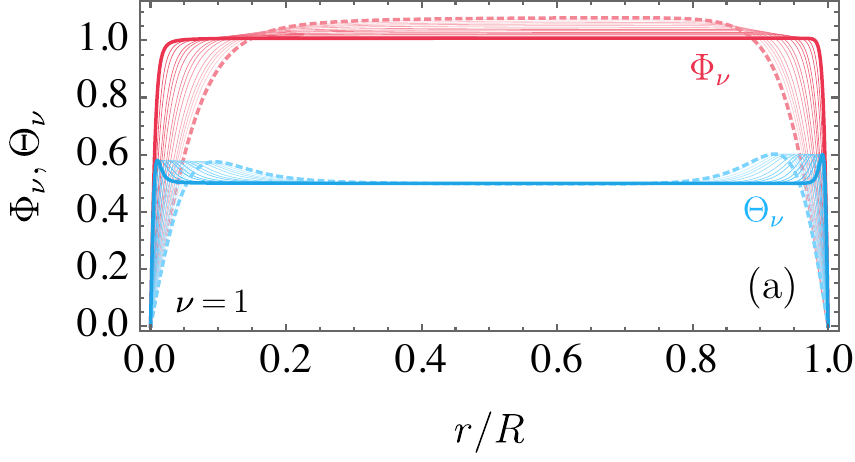}\linebreak{}
\par\end{centering}
\begin{centering}
\includegraphics{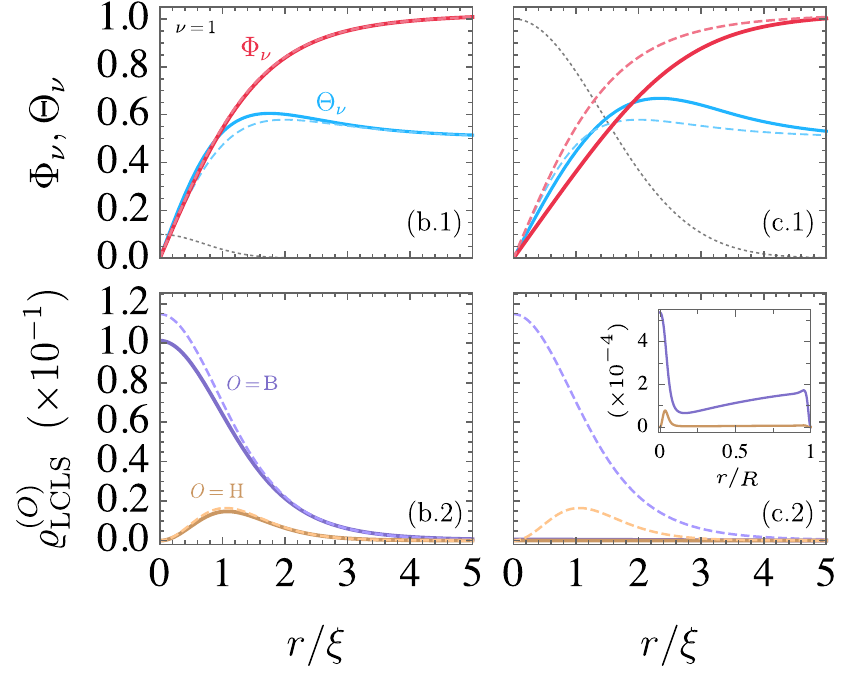}
\par\end{centering}
\centering{}\caption{Panel (a) shows plots of the wave functions of the BEC $\Phi_{\nu}$
(red) and of its adjoint $\Theta_{\nu}$ (blue) for $\nu=1$ in absence
of a pinning potential, for system sizes ranging between $R=20$ (dashed)
and $R=200$ (solid), across the entire radial dimension of the system.
The region near the vortex core is shown in subpanels (b.1\textendash{}2) for
a configuration of the pinning potential $(\epsilon,w)=(0.1,1)$ and
in subpanels (c.1\textendash{}2) for $(\epsilon,w)=(1,2)$. (b.1,c.1): wave functions
of the BEC $\Phi_{\nu}$ (red) and of its adjoint $\Theta_{\nu}$
(blue), where black dotted lines plot the profile of the pinning potential
$V_{\text{p}}$ for each of the configurations. (b.2,c.2): spatial
densities of the LCLS $\varrho_{\text{LCLS}}^{(O)}$ for $O=\text{B}$
(solid purple) and $O=\text{H}$ (solid orange); the inset in subpanel (c.2)
shows the spatial densities at a smaller scale across the
entire radial dimension, since they become invisible
at the scale of the plot and delocalized from the vortex core (as explained in Sec.~\ref{subsec:Effect-of-the-pinning-potential});
for comparison, dashed lines plot the respective spatial densities
in absence of a pinning potential. \label{fig:fig-1}}
\end{figure}

Combining Eq.~\eqref{eq:Form_GP-eq} and~\eqref{eq:Vortex_single-vortex-sol}
yields the reduced radial equation
\begin{equation}
\left(h+\frac{\nu^{2}}{r^{2}}-\mu+\phi_{\nu}^{2}(r)\right)\phi_{\nu}(r)=0\,.\label{eq:Methods_GPE-reduced-vortex}
\end{equation}
We generate numerical solutions of~\eqref{eq:Methods_GPE-reduced-vortex}
for $\nu=1$ using a combination of imaginary-time evolution and an
$R$-asymptotic approximation, as outlined in Appendix~\ref{sec:Computation-of-the};
results are presented in Fig.~\ref{fig:fig-1}(a). We then use
these to obtain numerical solutions of Eq.~\eqref{eq:Vortex_BdG-eq-reduced}
using a discretization-based solver.

Single-charge (i.e., $\bigl|\nu\bigr|=1$) vortices are \emph{dynamically
stable}, meaning that the associated Bogoliubov operator~\eqref{eq:Form_BdG-op}
possesses only real eigenvalues. However, due to a negative energy
of the LCLS, they are \emph{energetically unstable}, meaning that
the spectrum of $\hat{H}_{\text{B}}$, Eq.~\eqref{eq:Form_BdG-diag}, has a negative eigenvalue and,
therefore, that the mean-field~\eqref{eq:Methods_GPE-reduced-vortex}
is energetically unstable. The existence of the LCLS has long been
recognized and known to trigger the vortex's spiraling-out
motion~\cite{Dalfovo:1996aa,Rokhsar:1997aa,Isoshima:1997aa,Dodd:1997aa,Fetter:1998aa,Isoshima:1999ac,Fedichev:1999aa,svidzinsky_2000,Virtanen:2001ab}.
In Fig.~\ref{fig:fig-1} we show features of the
BEC wave function and its adjoint near the vortex core, as well as the Bogoliubov and hydrodynamic
spatial densities of the LCLS.

\subsection{Total densities of states}

Figure~\ref{fig:fig-2} shows plots of the Bogoliubov and the hydrodynamic
DOS for multiple values of $R$ and fixed $(\epsilon,w)$. We have
chosen to represent the Dirac delta function in Eqs.~\eqref{eq:LDOS_Bogol-ldos}
by a Lorentz distribution, 
\[
\delta(\omega)\sim\frac{1}{\pi}\frac{\delta\omega}{\omega^{2}+\delta\omega^{2}}\,,
\]
with the width $\delta\omega=\mathcal{O}(R^{-2})$; this width is
of the energy scale of a single particle in a rigid wall potential
of size $R$ and yields the Dirac delta in the limit $R\rightarrow\infty$.
We see that, apart from noise, the numerical results are in good
agreement with the analytical results in Eqs.~\eqref{eq:LDOS_continuum-Bogol-dos}
and~\eqref{eq:LDOS_continuum-Hydro-dos} for the homogenous BEC.

Both $D_{B}^{(\text{B})}$ and $D_{B}^{(\text{H})}$
deviate from the numerical result only in the low-energy region of
the spectrum, around $\omega\sim0$, as shown in the insets of Figs.~\ref{fig:fig-2}.
This deviation is twofold: i) the low-lying hydrodynamic DOS is dominated
by a peak at energy $\omega_{-1,0}$ of the LCLS (or remnant kelvon
mode) {[}inset of Fig.~\ref{fig:fig-2}(b){]}; ii) the Bogoliubov DOS shows that the energy $\omega_{0,0}=\omega_{0,0}(R)$, the energy
of the first \emph{non-kelvonic} excitation, appears to become larger
than the typical inter-level spacing with increasing $R$ {[}highlighted in the inset
of Fig.~\ref{fig:fig-2}(a){]}. Additionally, the presence of the
NG mode (especially notable in the hydrodynamic DOS) is not inconsistent
with the Bogoliubov spectrum $\omega_{B}(\mathbf{k})$, which
accounts only for density excitations. (Although excitations of the Bogoliubov spectrum entail phase excitations, the NG mode is an excitation of the phase \emph{exclusively}.)

\subsection{Angular momentum-resolved densities of states}

\subsubsection{AM-DOS}

\begin{figure}
\begin{centering}
\includegraphics{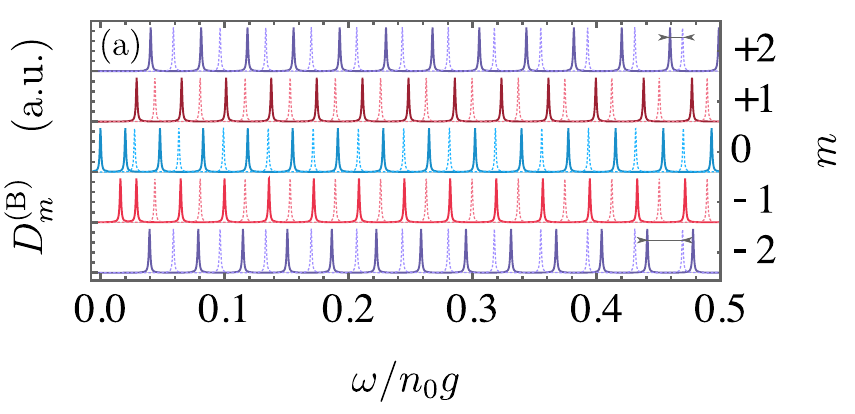}\linebreak{}
\par\end{centering}
\centering{}\includegraphics{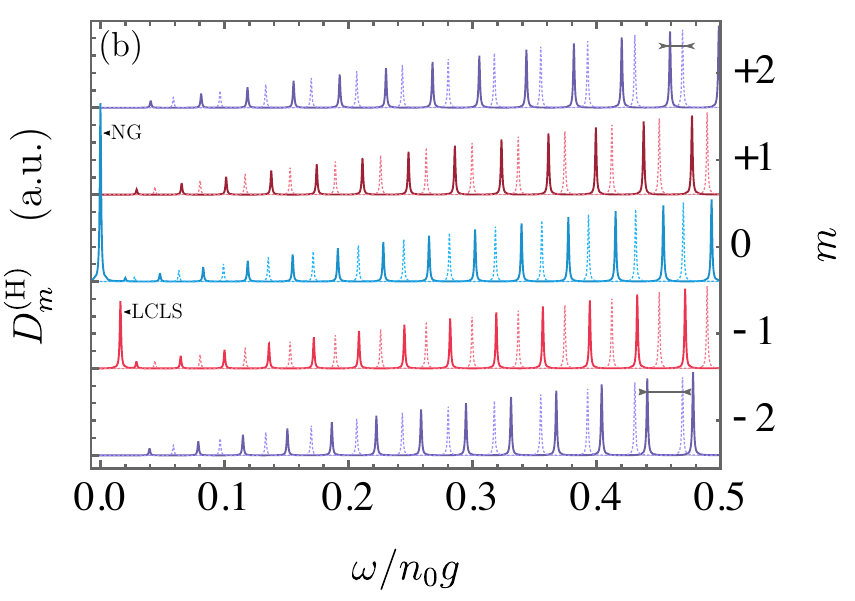}\caption{Plots of the (a) Bogoliubov and (b) hydrodynamic AM-DOS for angular
momenta $m=0,\pm1,\pm2$ at system size $R=125$ and pinning $(\epsilon,w)=(0.1,1)$.
Dotted lines are the corresponding AM-DOS for the fully homogeneous
BEC in Eqs.~\eqref{eq:LDOS_continuum-Bogol-dos} and~\eqref{eq:LDOS_continuum-Hydro-dos}. Panel (b): peak intensities are proportional to the hydrodynamic
weights $c_{m,n}^{(\text{H})}$; within the energy range plotted,
the largest is $c_{\text{NG}}^{(\text{H})}\approx0.5$ of the Nambu-Goldstone
mode. In the plots for $m=\pm2$ we have marked energy differences
with respect to the homogeneous BEC to highlight the lifting of angular-momentum
degeneracy by the vortex. \label{fig:fig-3}}
\end{figure}

Figure~\ref{fig:fig-3} shows plots of the AM-DOS for selected values
of $(\epsilon,w)$ and $R$ and angular momenta $m=0,\pm1,\pm2$.
Notable features of the hydrodynamic DOS are reproduced here: the
NG mode, at $m=0$, and the LCLS, at $m=-1$, tower over all other
low-lying states.

\begin{figure}
\begin{centering}
\includegraphics{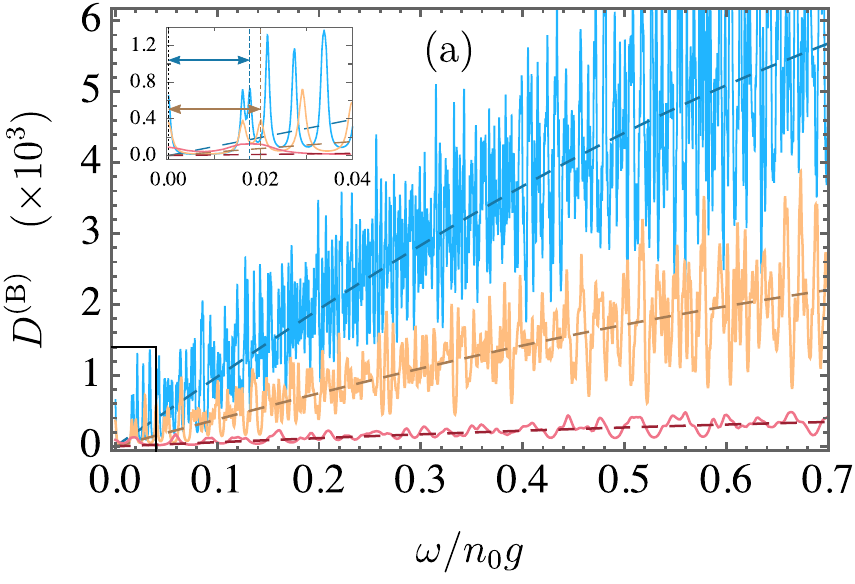}\linebreak{}
\par\end{centering}
\centering{}\includegraphics{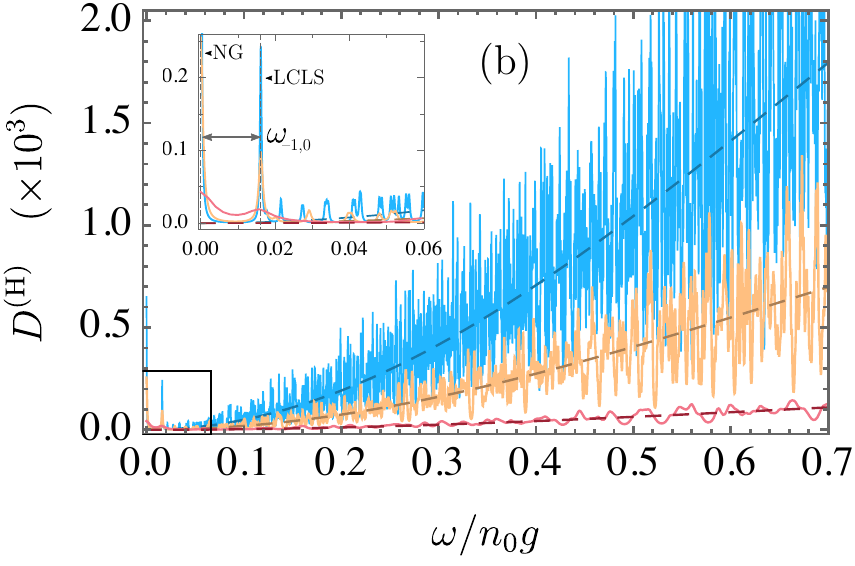}\caption{Plots of the (a) Bogoliubov and (b) hydrodynamic DOS (solid), compared with
the plots of Eqs.~\eqref{eq:LDOS_continuum-Bogol-dos} and~\eqref{eq:LDOS_continuum-Hydro-dos}
(dashed), for system sizes $R=50$ (red), $R=125$ (orange) and $R=200$
(blue) (appearing in ascending order in the plots), and pinning $(\epsilon,w)=(0.1,1)$.
Distinct scales in the vertical axes show that $D^{(\text{B})}$ bounds
$D^{(\text{H})}$. Insets zoom into the low-energy features in the
boxed regions of the respective plots: (a) the highlighted energies
are the energy of the first density (non-kelvonic) excitation $\omega_{0,0}$
for the two system sizes $R=125$ and $R=200$; (b) peaks belonging
to the NG mode and LCLS are labelled; the energy $\omega_{-1,0}$
is the energy of the LCLS at the selected configuration of the pinning
potential. The notation of the energy levels follows Eq.~\eqref{eq:Vortex_BdG-eq-reduced}.\label{fig:fig-2}}
\end{figure}

There is a visible growth of the hydrodynamic weight, Eq.~\eqref{eq:LDOS_Hydro-wght} (the height of
the peaks in each $D_{m}^{(\text{H})}$), with energy that ties in
with the known breakdown of the hydrodynamic approximation beyond
low energies~\cite{Dalfovo:1999aa}: it signals a departure
from a collective sound-wave (phononic) picture of excitations, wherein
the spectrum is $\sim k$, to a single-particle (atomic) one, with
the spectrum $\sim k^{2}$. The transition from the phononic to the
atomic picture is accompanied by a decrease in the magnitude of the
hole-component $v_{\lambda}$ relative to the particle-component $u_{\lambda}$
of the excitation~\cite{Dalfovo:1999aa}. Thus, in this sense, the
hydrodynamic weight $c^{(\text{H})}$, Eq.~\eqref{eq:LDOS_Hydro-wght},
is a measure of the particle-hole imbalance of a bosonic state. We
explain this observation by noting that the hydrodynamic spatial
density $\varrho_{\lambda}^{(\text{H})}$, Eq.~\eqref{eq:LDOS_hydro-spat-dens},
amounts to an \emph{interference pattern} between the amplitudes of
the particle and hole components $u_{\lambda}$ and $v_{\lambda}$;
this interference is essentially destructive, since $\Phi_{0}v_{\lambda}$
has only a phase $e^{\text{i}\pi}$ relative to $\overline{\Phi}_{0}u_{\lambda}$.
Thus, the discrepancy between the magnitudes of $u_{\lambda}$ and
$v_{\lambda}$ determines the intensity of the interference pattern
$\varrho_{\lambda}^{(\text{H})}$ and then $c_{\lambda}^{(\text{H})}$
, being its integral, functions as a \emph{global }measure of the
particle-hole \emph{imbalance} of the mode $X_{\lambda}$.

We note, moreover, that the presence of the vortex is known to lift
the angular momentum-degenerate excitations of an otherwise homogeneous
BEC~\cite{Svidzinsky:1998aa}, a feature that we highlight in
Fig.~\ref{fig:fig-3}.

\subsubsection{AM-LDOS}

\begin{figure}[t]
\centering{}\includegraphics{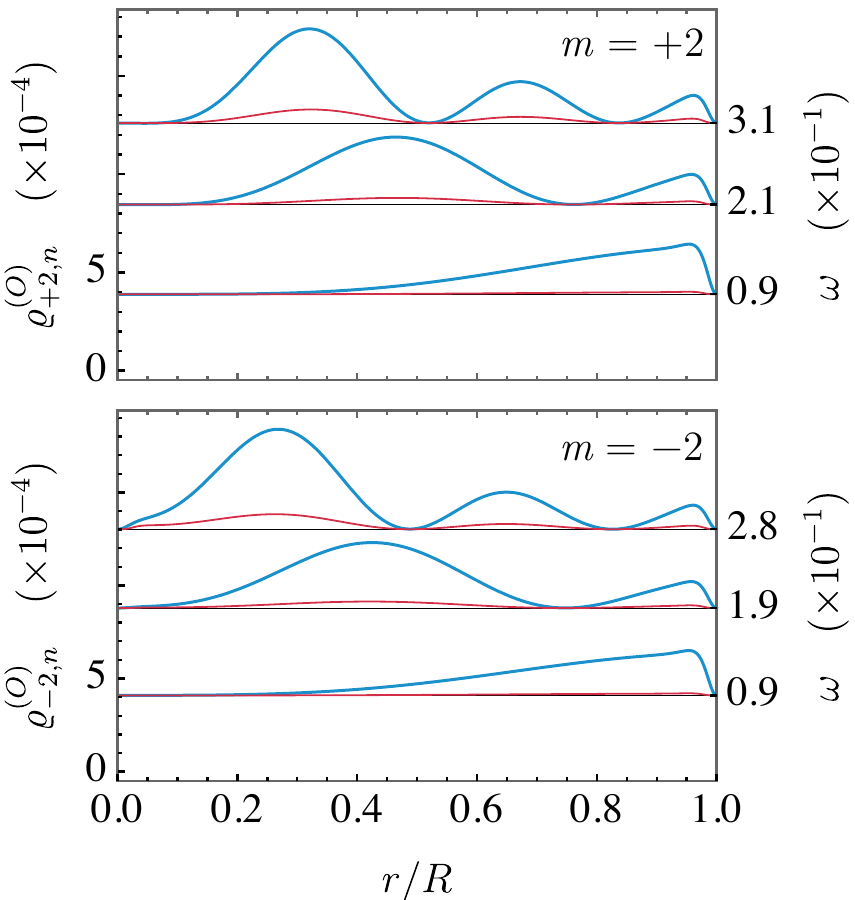}\caption{Plots of the Bogoliubov (blue, $O=\text{B}$) and hydrodynamic (red,
$O=\text{H}$) spatial densities for the first few states of angular
momentum $m=0,\pm1,\pm2$ at system size $R=50$ and pinning $(\epsilon,w)=(0.1,1)$,
representing the AM-LDOS at each value $\omega=\omega_{m,n}$ along
the right-vertical axes. Magnitudes of spatial densities can be inferred
from the scale on the left-vertical axis of each panel; each Bogoliubov
spatial density has locally a larger magnitude than its hydrodynamic
counterpart. $m=\pm2$: densities are nearly identical except at $r\approx0$
(more visible with increasing energies). $m=+1$: the Bogoliubov density
is hole-like (negative) at $r\approx0$. $m=0$: the lowest state
(NG mode) is plotted in distinct colors\textemdash purple for $O=\text{B}$,
orange for $O=\text{H}$; unlike most other states, the hydrodynamic
spatial density is comparable with its Bogoliubov counterpart. $m=-1$:
the lowest state (LCLS) is plotted in likewise distinct colors and
in dotted lines, shown here for comparison\textemdash a clearer picture
displayed in Fig.~\ref{fig:fig-1}(b.2); here, the Bogoliubov density
is purposefully shown to exceed the plot range.\label{fig:fig-4}}
\end{figure}

Contrarily to the DOS, the AM-DOS of a finite system are sparse (compare
Figs.~\ref{fig:fig-2} and~\ref{fig:fig-3}), that is, states are
separable within each angular momentum sector. It follows that
the spatial dimension of the AM-LDOS can be represented faithfully
in terms of individual spatial densities $\varrho_{m,n}^{(O)}$, Eqs.~\eqref{eq:LDOS_Bogol-spat-dens}
and~\eqref{eq:LDOS_hydro-spat-dens}, alone. These are displayed
in Fig.~\ref{fig:fig-4}. In particular, the Bogoliubov density of
the LCLS is shown to be orders-of-magnitude larger at the core than
other low-lying, core-localized states.

This happens because $\varrho^{(\text{B})}$ in Eq.~\eqref{eq:LDOS_Bogol-spat-dens}
is sign-indefinite, so that a bosonic state can be locally particle-like
(hole-like) in regions of positive (negative) sign; accordingly, the
state is locally characterized by an accumulation (depletion) of atoms
proportional to its magnitude. Complementarily to the hydrodynamic
weight $c^{(\text{H})}$, which acts as a global measure, the Bogoliubov
spatial density $\varrho^{(\text{B})}$ acts as a \textit{local} measure
of particle-hole \emph{character} of a state. The LCLS is, therefore,
largely more particle-like ($\vert u_{\text{LCLS}}\vert\gg\vert v_{\text{LCLS}}\vert$)
than most other low-lying states~\cite{Rokhsar:1997aa}, whence
its hydrodynamic weight derives exceptional magnitude. The one exception
is the NG mode, which is likewise particle-like but delocalized: its
Bogoliubov AM-LDOS is similar in magnitude to that of the first non-kelvonic
excitation, for instance, but the hydrodynamic AM-DOS of the latter
is vanishing\textemdash in fact, the hydrodynamic weight of the first
is negligible while that of the NG surpasses the LCLS (see Fig.~\ref{fig:fig-3}).
As energy increases, the magnitudes of the Bogoliubov and hydrodynamic
densities become generically comparable, as modes become progressively
more particle-like. These features are patent in Fig.~\ref{fig:fig-4}.

\begin{figure}[t]
\centering{}\includegraphics{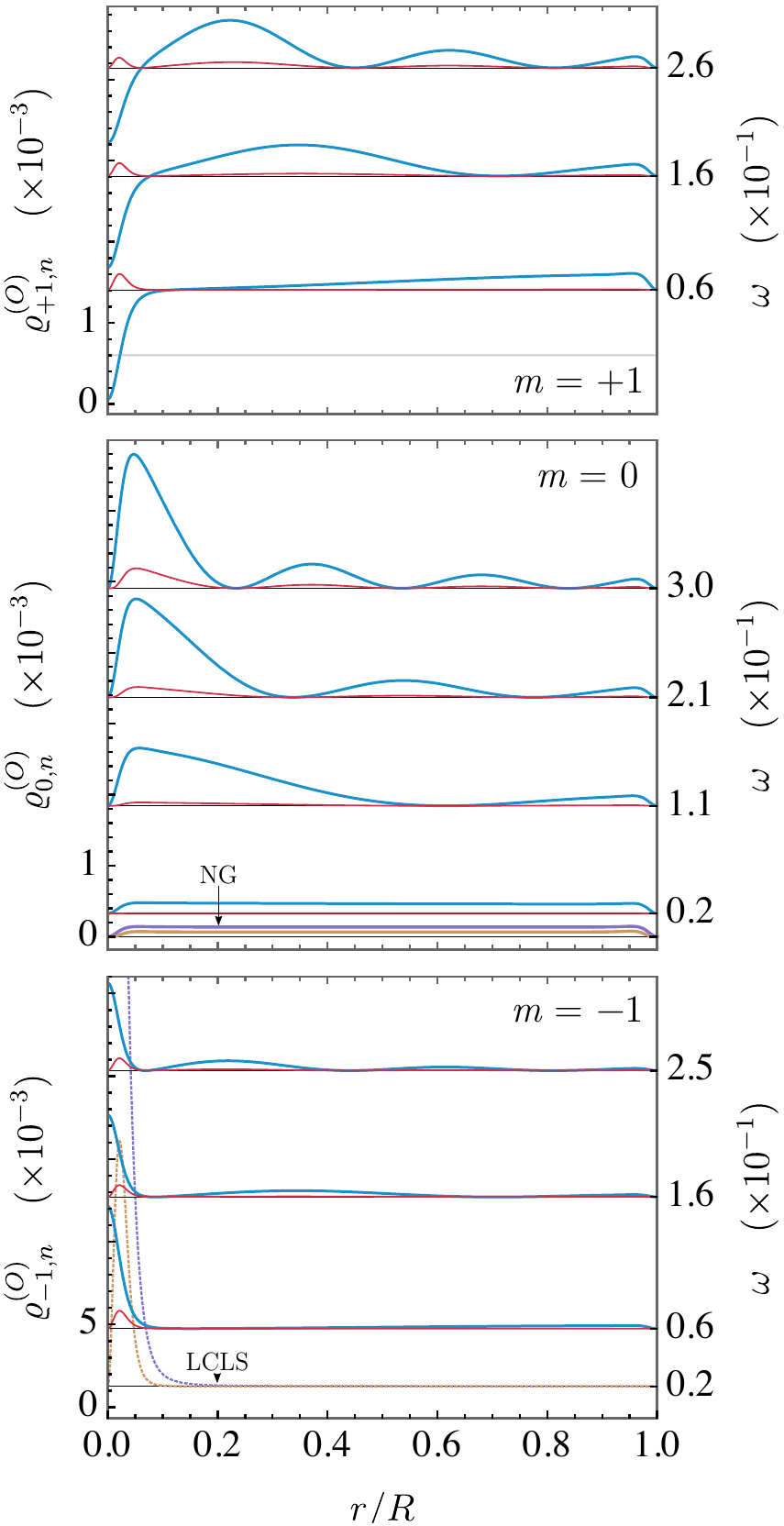}
\end{figure}

\subsection{Features of low-lying states}

\subsubsection{Effect of the pinning potential\label{subsec:Effect-of-the-pinning-potential}}

\begin{figure}
\centering{}\includegraphics{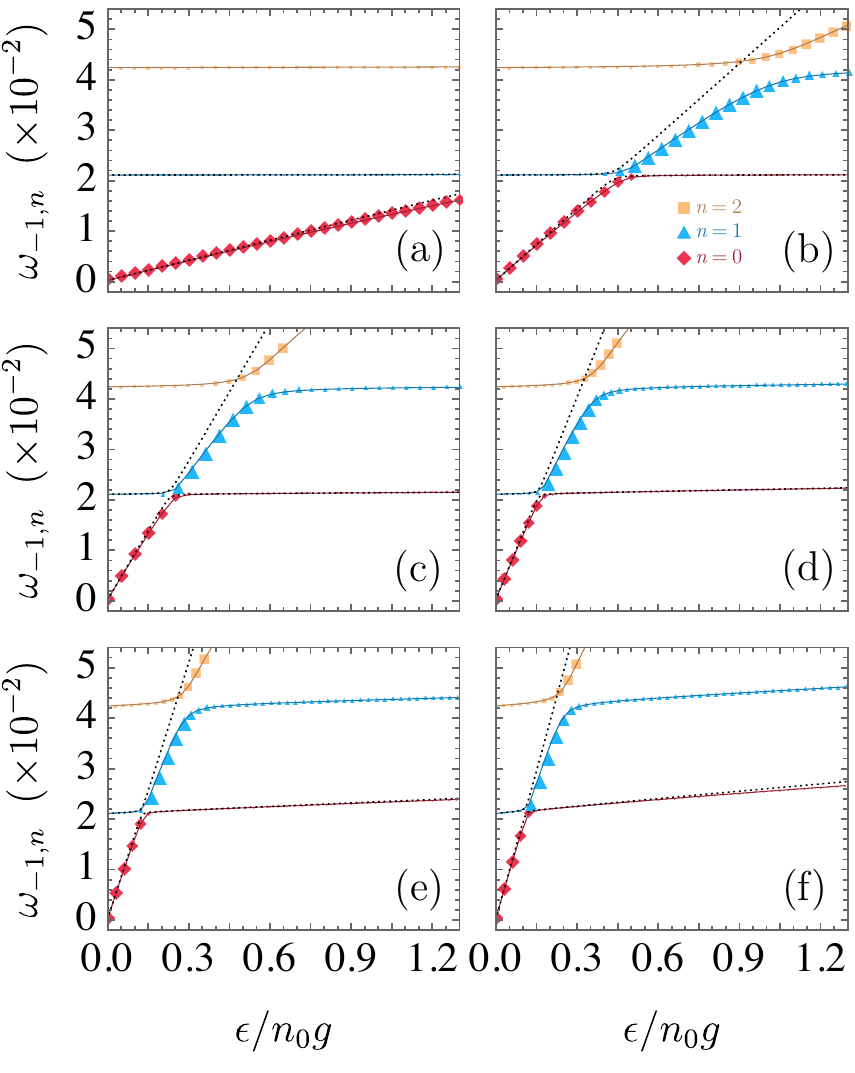}\caption{Plots of eigenvalues $\omega_{m,n}$ of the modes $n=0,1,2$ of angular
momentum $m=-1$ as a function of the maximum optical potential $\epsilon$,
for $R=200$ and beam waist (a) $w=0.2$, (b) $w=0.4$, (c) $w=0.6$,
(d) $w=0.8$, (e) $w=1.0$ and (f) $w=1.2$; the size of the plot
markers is proportional to the hydrodynamic weight, Eq.~\eqref{eq:LDOS_Hydro-wght},
of each state, for each value of $\epsilon$ (scaled logarithmically
for comparison). Black dotted lines show the results of the minimal
model described in the text.\label{fig:fig-5}}
\end{figure}

The following results show how the LCLS is strongly dependent on the
pinning potential, while other modes have only a negligible (and indirect)
dependence. We begin by noting that the pinning energy competes
against the dominant centrifugal barrier $(m+\nu\sigma_{3})^{2}/r^{2}$
at the vortex core $r\lesssim1$ in Eq.~\eqref{eq:Vortex_BdG-reduced}.
A mode $(m,n)$ is, thus, insensitive to the pinning as long as it
cannot penetrate the centrifugal barrier: there may be a non-negligible
dependence on the pinning potential only in case $\omega_{m,n}\gtrsim(m\pm\nu)^{2}$.
This observation ensures that we will find a negligible effect for
all low-energy modes except at angular momenta $m=\pm1$; we have
found the numerical evidence to support this observation, and so we
focus our discussion on the case $m=-1$, for definiteness. In Fig.~\ref{fig:fig-5},
we depict the eigenvalues and hydrodynamic weights $c^{(\text{H})}$,
Eq.~\eqref{eq:LDOS_Hydro-wght} (shown as the size of plot markers),
at low energies, as functions of the pinning parameters. The strong
dependence of the LCLS on the pinning potential is visible and, furthermore,
we see that the $c^{(\text{H})}$ cross over at avoided level crossings,
suggesting that the LCLS enters the energy region of (non-kelvonic)
density excitations. In order to clarify these features, we derived
the minimal model described next.

\begin{figure}
\centering{}\includegraphics{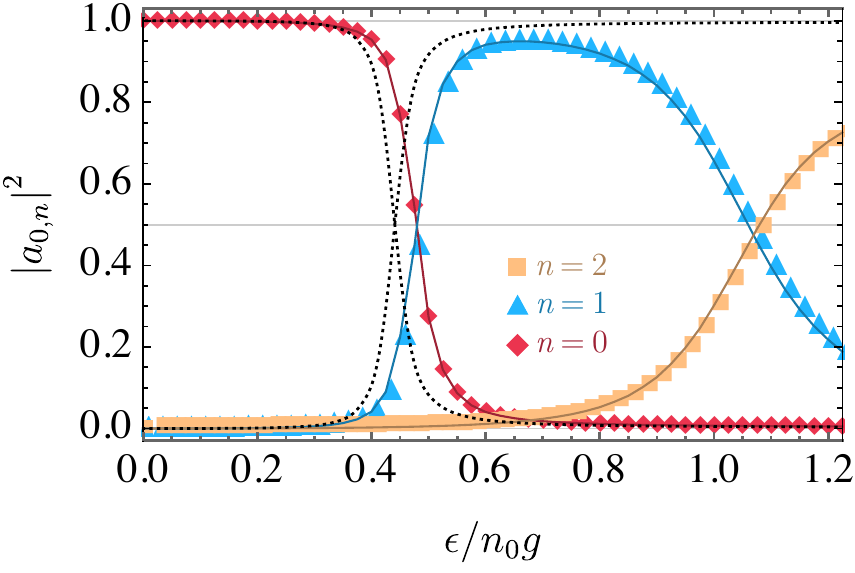}\caption{Plots of overlaps $\left|a_{0,n}\right|^{2}=\bigl|\bigl\langle X_{0}^{(0)},X_{n}\bigr\rangle\bigr|^{2}$
of modes $n=0,1,2$ of angular momentum $m=-1$ with the $n=0$ mode
of the pinning-less system, as a function of the maximum optical potential
$\epsilon$ for system size $R=200$ and beam waist $w=0.4$ {[}same as Fig.~\ref{fig:fig-5}(b){]}.
Black dotted lines show the results of the minimal model described in
the text. The curve for $n=3$ is not included, though it is visible
that $a_{0,3}$ becomes non-negligible within this range.\label{fig:fig-6}}
\end{figure}

We consider the reduced Bogoliubov Eq.~\eqref{eq:Vortex_BdG-eq-reduced}
with the Hamiltonian rewritten as $\mathcal{H}=\mathcal{H}^{(0)}+\Delta\mathcal{H}$,
where
\begin{equation}
\Delta\mathcal{H}=\sigma_{3}V_{\text{p}}+\sigma_{3}\delta_{\text{BEC}}\,,\label{eq:Tun_pin-pot_pert}
\end{equation}
with
\begin{equation}
\delta_{\text{BEC}}=\left(\begin{array}{cc}
-\delta\mu+2\delta\phi^{2} & \delta\phi^{2}\\
\delta\phi^{2} & -\delta\mu+2\delta\phi^{2}
\end{array}\right)\,,\label{eq:Tun_deform-BEC}
\end{equation}
(we omit subscripts $m=-1$ and $\nu=1$ within this section), where
$\delta\mu=\mu-\mu_{0}$ and $\delta\phi^{2}=\phi^{2}-\phi_{0}^{2}$;
in the absence of a pinning potential, $\phi_{0}$ and $\mu_{0}$
are solutions of the GP Eq.~\eqref{eq:Form_GP-eq} and $\mathcal{H}^{(0)}$
is the reduced Hamiltonian. Therefore, the term $\delta_{\text{BEC}}$
is understood as a potential energy due to the \emph{deformation }of
the BEC caused by the applied pinning potential.

The first level avoidance involves the $m=-1$ modes $X_{n}^{(0)}$,
for $n=0,1$, solutions of $\mathcal{H}^{(0)}X_{n}^{(0)}=\omega_{n}^{(0)}X_{n}^{(0)}$.
Thus, we expand the eigenstates of~\eqref{eq:Vortex_BdG-eq-reduced}
in this subspace, i.e,
\begin{equation}
X(r)=a_{0}X_{0}^{(0)}(r)+a_{1}X_{1}^{(0)}(r)\,.\label{eq:Tun_min-model-sols}
\end{equation}
Taking the bilinear product $\bigl\langle X_{i}^{(0)},\mathcal{H}X\bigr\rangle$,
for $i=0,1$, yields the algebraic equation
\[
\left(\begin{array}{cc}
\omega_{0}^{(0)}+\Delta_{00} & \Delta_{01}\\
\Delta_{10} & \omega_{1}^{(0)}+\Delta_{11}
\end{array}\right)\left(\begin{array}{c}
a_{0}\\
a_{1}
\end{array}\right)=\omega\left(\begin{array}{c}
a_{0}\\
a_{1}
\end{array}\right)\,,
\]
where $\Delta_{ij}=\Delta_{ji}=\bigl\langle X_{i}^{(0)},\Delta\mathcal{H}X_{j}^{(0)}\bigr\rangle$,
with the eigenvalues
\begin{equation}
\omega_{n=0,1}=\tilde{\omega}+\Delta_{+}\mp\sqrt{\left(\Omega-\Delta_{-}\right)^{2}+\left|\Delta_{01}\right|^{2}}\,,\label{eq:Tun_min-model-eigvals}
\end{equation}
where $\tilde{\omega}=(\omega_{1}^{(0)}+\omega_{0}^{(0)})/2$, the
half-gap $\Omega=(\omega_{1}^{(0)}-\omega_{0}^{(0)})/2$ and $\Delta_{\pm}=\left(\Delta_{00}\pm\Delta_{11}\right)/2$;
the minus- (plus-) signed branch in Eq.~\eqref{eq:Tun_min-model-eigvals}
is the $n=0$ ($n=1$) solution. The comparisons of Eqs.~\eqref{eq:Tun_min-model-eigvals}
with the numerical results in Fig.\textcolor{teal}{~}\ref{fig:fig-5}
reveal the qualitative agreement of the minimal model; the quantitative
inaccuracy results simply from the truncated subspace in Eq.~\eqref{eq:Tun_min-model-sols}
and is of no bearing to the following analysis.

Further comparing Eq.~\eqref{eq:Tun_min-model-eigvals} with numerics
in Fig.\textcolor{teal}{~}\ref{fig:fig-5}, we observe $\Delta_{00}$
to be much larger than $\Delta_{01}$ and $\Delta_{11}$; indeed,
we found $\Delta_{01}\approx10^{-2}\Delta_{00}$ while $10^{-2}<\Delta_{11}/\Delta_{01}\lesssim1$
across the sampled values of $(\epsilon,w)$. To clarify these disparities,
we consider the quantity
\[
\eta_{ij}=\frac{\bigl\langle X_{i}^{(0)},\sigma_{3}\delta_{\text{BEC}}X_{j}^{(0)}\bigr\rangle}{\bigl\langle X_{i}^{(0)},\sigma_{3}V_{\text{p}}X_{j}^{(0)}\bigr\rangle}\,,
\]
that is, the ratio of the contributions to the $\Delta_{ij}$, Eq.~\eqref{eq:Tun_pin-pot_pert}:
the term $V_{\text{p}}$ is the potential energy due the pinning potential;
the term $\delta_{\text{BEC}}$, Eq.~\eqref{eq:Tun_deform-BEC},
is the potential energy due the deformation of the BEC caused by the
pinning potential. (Note that both terms are effects of the application
of the pinning potential on the system, but that $V_{\text{p}}$ is
the \emph{direct} effect while $\delta_{\text{BEC}}$ is an \emph{indirect}
effect on its modes, in particular.) We found that $\eta_{11}>0$
with $0.1<\eta_{11}<20$ and, for $i=0,1$, $\eta_{0i}<0$ (negative
due to the deformation term) with $10^{-2}<\left|\eta_{0i}\right|<0.5$,
increasing with $w$ in all cases\textemdash that is, the term $V_{\text{p}}$
dominates over $\delta_{\text{BEC}}$ for both $\Delta_{00}$ and
$\Delta_{01}$ and vice-versa for $\Delta_{11}$. This shows, since
$\Delta_{00}\gg\Delta_{01}\gg\Delta_{11}$, that the LCLS is strongly
dependent on the pinning potential \emph{directly}, due to its exceptionally
large amplitude at the vortex core; other modes are, at most, negligibly
dependent on the pinning potential and \emph{indirectly} so via the
deformation of the BEC, as shown by $\Delta_{11}$.

\begin{figure}
\centering{}\includegraphics{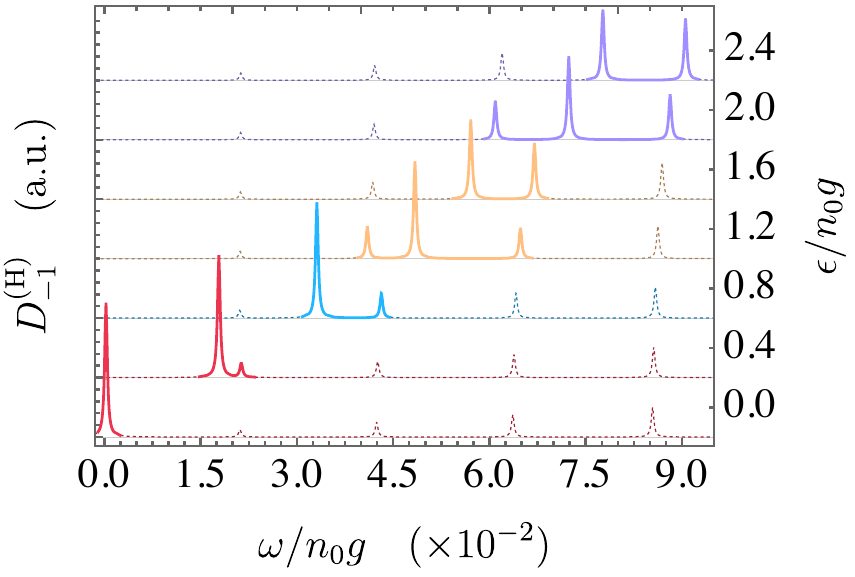}\caption{Hydrodynamic AM-DOS of low-energy modes of angular momentum $m=-1$ as a function of the maximum optical potential $\epsilon$ (right-vertical axis) for system size $R=200$ and beam waist $w=0.4$ [same as Fig.~\ref{fig:fig-5}(b)]. Solid lines highlight modes that have a noticeably increased hydrodynamic weight $c^{(\text{H})}$, for each value of $\epsilon$, and colours indicate the mode having the largest hydrodynamic weight with red ($n=0$), blue ($n=1$), orange ($n=2$) and violet ($n=3$). The fact that the energy of the LCLS is positive in the absence of a pinning potential ($\epsilon=0$) is addressed in Sec.~\ref{subsec:Scaling-and}. \label{fig:fig-7}}
\end{figure}

We analyze the crossover of the $c^{(\text{H})}$ by considering the
probability of observing the pinning potential-free LCLS (i.e., the
state $X_{0}^{(0)}$) given a state $X_{n}$, that is, the overlap
$\left|a_{0,n}\right|^{2}=\bigl|\bigl\langle X_{0}^{(0)},X_{n}\bigr\rangle\bigr|^{2}$,
plotted in Fig\textcolor{teal}{.~}\ref{fig:fig-6}. This figure of
merit differentiates the states that possess a large, particle-like
density at the vortex core for each configuration of the pinning potential.
In particular, we can conclude that the LCLS (i.e., the mode with
quantum numbers $m=-1$ and $n=0$) eventually loses the characteristic
hydrodynamic weight resulting from the large, core-localized
density, as this becomes, due to the pinning potential, a feature of
higher energy states. Hence the negligible magnitude of the $\varrho_{\text{LCLS}}^{(O)}$
seen in Fig.~\ref{fig:fig-1}(c.2) relative to Fig.~\ref{fig:fig-1}(b.2).
Moreover, we notice that the crossover $a_{0,0}$ with $a_{0,1}$
is relatively steep while $a_{0,1}$ with $a_{0,2}$ is markedly smoother
and can be seen to take place at a value $\left|a_{0,2}\right|^{2}<0.5$.
This suggests that, as the intensity of the pinning potential increases
(and with it the energy of a density at the vortex core), the large \begin{figure}[H]
\begin{centering}
\includegraphics{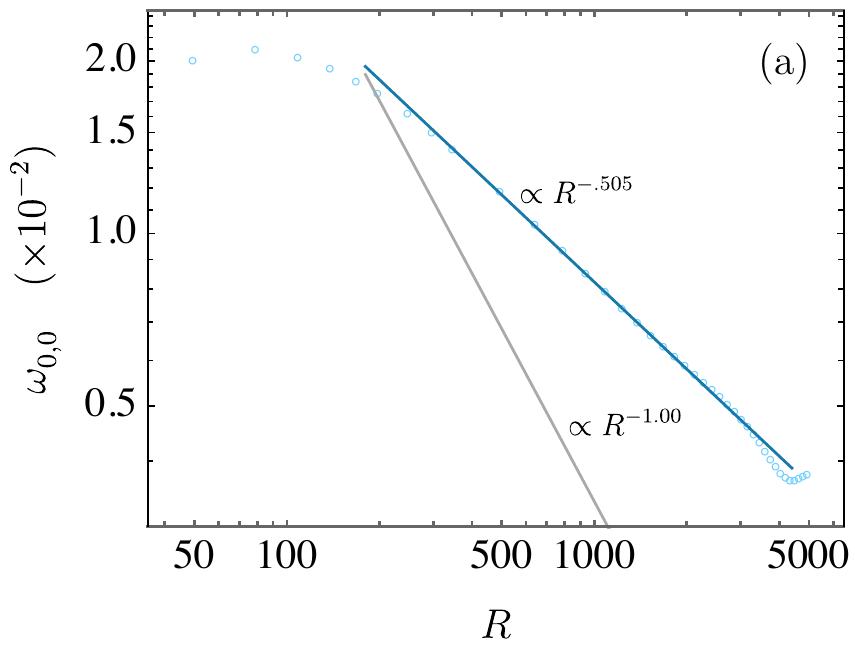}\linebreak{}
\par\end{centering}
\centering{}\includegraphics{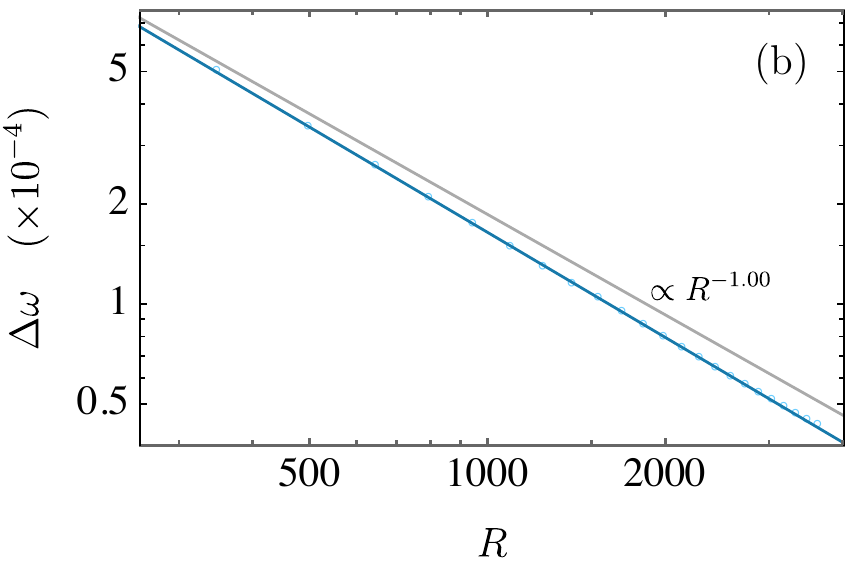}\caption{Log-log plots of (a) $\omega_{0,0}$, the energy of the first non-kelvonic
excitation without a pinning potential, and (b) $\Delta\omega$, the
mean inter-level spacing in energy between states above $\omega_{0,0}$,
as functions of $R$; light blue circles are the numerical data and
the darker blue lines are the linear fits performed in the scaling
region used to obtain the exponents $\alpha=0.505\pm0.002$ for $\omega_{0,0}^{(0)}$
and $\beta=1.05\pm0.03$ for $\Delta\omega$; the latter we obtained
from a fixed sample including the first 250 states above $\omega_{0,0}$
and, in the fitting, we considered the standard error of the mean.
Data breaks away from the scaling region at very large $R$ plausibly
due to numerical errors, as local features become too small for at
the defined numerical precision. Gray lines in each panel are the
corresponding quantity obtained from the homogeneous system used to
derive Eqs.~\eqref{eq:LDOS_continuum-Bogol-dos} and~\eqref{eq:LDOS_continuum-Hydro-dos},
whose energy levels are given explicitly in terms of the homogeneous
spectrum by $\omega_{m,n}=\omega_{B}(k_{m,n})$, $k_{m,n}=j_{m,n+1}/R$
where $j_{m,n+1}$ is the $(n+1)$-th zero of the Bessel function
$J_{m}$; deviations from the $R^{-1}$ scaling are one order of magnitude
below significant digits and covered by error margins. \label{fig:fig-8}}
\end{figure} hydrodynamic weight characteristic of the LCLS (for a weak pinning
potential) becomes spread out across a number of modes instead of concentrated in a single one, and, so, there will be a number of modes
with increased hydrodynamic weights instead of a single dominant one.
Thus, we can think of the large hydrodynamic weight, initially concentrated
in the LCLS, as becoming diluted under a sufficiently intense pinning
potential. This effect is shown in terms of the hydrodynamic AM-DOS in Fig.~\ref{fig:fig-7}.

\subsubsection{Scaling and $R$-dependence\label{subsec:Scaling-and}}

We carried out scaling analyses concerning the discrepancy
between the gap (that is, the energy of the lowest non-kelvonic excitation
$\omega_{0,0}$) and the mean inter-level spacing $\Delta\omega$,
as apparent in the inset of Fig.~\ref{fig:fig-2}(a). Thus, we obtained
information on the appropriate low-energy, continuum description of
density excitations; results are shown in Fig.~\ref{fig:fig-8}.
Indeed, we find distinct scalings between the energy of the first
density excitation and the mean inter-level spacing, indicative of
a gap that scales anomalously as $\mathcal{O}(R^{-1/2})$. For comparison, we show
the prediction for a homogeneous BEC (given by the Bogoliubov spectrum
$\omega_{B}$) to scale as $R^{-1}$; this gap is a trivial
finite-size effect, as it scales equally to the mean inter-level spacing
and approaches a massless spectrum.
Notice that the deviation in the exponent of the mean inter-level
spacing from that of the homogeneous spectrum, though small, is not
accounted for by the error margin. Considering that the group velocity must vanish at the gap, this deviation is suggestive of a non-analytic, momentum-dependent correction to the linear dispersion.

Finally, it shows in Fig.~\eqref{fig:fig-6} that the energy level
of the LCLS does not cross zero at the presented system size, that
is, that it does not represent an energetic instability. Indeed, we have
found that this mode stabilizes spontaneously for a system size $R\gtrsim73$,
in qualitative agreement with Ref.~\cite{Rokhsar:1997aa}. (Eventual
quantitative discrepancies are attributed to the fact that a quasi-2D
BEC produced by a box potential is considered here.)

\section{Discussion and Conclusions\label{sec:Discussion-and-Conclusions}}

We have acquired new insights into the spectral properties of a quantum vortex in a quasi-2D BEC by analyzing the symplectic and hydrodynamic densities of states (DOS). Our discussion was particularly focused on two modes of the system. On the one hand, the lowest core-localized state (LCLS), which is the remnant kelvon mode surviving the dimensionality reduction imposed by the trap. On the other, the Nambu-Goldstone (NG) mode inherent to the BEC state. Most strikingly, we have shown that these modes have an exceptional capacity to interact with impurities embedded in the BEC. Therefore, they can have a sizeable influence on the dynamics of these impurities, or, more generally, heterogeneous atomic species. Regarding the NG mode, we propose that this effect may be observable in the dynamics of degenerate fermion-BEC mixtures as well as heterogeneous BEC mixtures~\cite{Belemuk:2006tu,Eto:2011tb,Indekeu:2018vu,Mithun:2021wi}. A proper account of the NG mode also becomes important to the polaron physics of the impurity-BEC system, as impurities can become exceptionally sensitive to the phase fluctuations of the BEC. Physically, this effect is rooted in the particular way excitations interact with an impurity, of which strength results from an interference pattern between their particle and hole components.

The LCLS will greatly affect impurities trapped in the core of the vortex, due to its localized character. It can stimulate transitions between quantum states of the impurity, be them orbital states or internal degrees of freedom, by virtue of the LCLS being an eigenstate of angular momentum. In turn, the LCLS is highly sensitive to the action of a pinning potential (this is the physical fundament of that mechanism of vortex stabilization~\cite{Isoshima:1999ac,Isoshima:1999ab}). It follows that the LCLS affords a novel mechanism of control over a vortex-trapped impurity at the quantum level: its energy can be tuned to a transition between two states of distinct angular momentum of the impurity, by means of the pinning potential, and stimulate the transition. As a matter of fact, there is no fundamental reason for this mechanism to be limited to the single-vortex configuration considered presently\textemdash we can expect local properties of the excitations to hold in different physical setups. For instance, in the Abrikosov lattice of a quasi-2D BEC, each vortex in the array is bound to possess a qualitatively similar remnant kelvon mode~\cite{Chevy:2003aa}. By allowing for such a degree of local, quantum-level control, this channel may be of practical use for a proposal based on vortex-trapped impurities as qubit units for quantum information processing~\cite{Shaukat:2017tc,Shaukat:2019aa,Braz:2020aa}.

Based on the symplectic DOS, we have inferred anomalies in the spectrum of excitations, namely, that the excitations of a large (but finite) BEC hosting a vortex are slightly (but non-vanishingly) gapped. This is anomalous with respect to the typical Bogoliubov spectrum, which is gapless and linear. One implication of a gapped dispersion is that the group velocity must vanish at the gap. Our scaling analysis of the mean inter-level spacing is consistent with this condition: it presents a non-negligible deviation from linearity, suggesting a non-analytic dependence in momentum. Physically, we attribute these results to the non-local, long-range profile of the vortex: its decays as $\sim1/r^{2}$ results in the $\log(R)$-divergent energy of the BEC and, notably, in a logarithmically-modified dispersion of its kelvons (in a three-dimensional BEC)~\cite{Pitaevskii:1961aa}. To our knowledge, however, the theory to support the long-wavelength dispersion of the in-plane, non-kelvonic density excitations is yet to be established. This may be important to the quantum treatment of vortex dynamics in a quasi-2D BEC~\cite{Simula:2020aa}.

\begin{acknowledgments}
The authors acknowledge the financial support of Funda\c{c}\~ao para a Ci\^encia e Tecnologia (FCT-Portugal) through Grant No. PD/BD/128625/2017, Contract No. CEECIND/00401/2018, Grant No. UID/CTM/04540/2019, Project PTDC/FIS-OUT/3882/2020, and Grant No. COVID/BD/151814/2021. H.T. further acknowledges the financial support from the Quantum Flagship Grant PhoQuS (Grant No. 820392) of the European Union. J.E.H.B. is grateful to Ana Vald\'es and Francisco Salces for ingenious input on experimental set-ups.
\end{acknowledgments}

\appendix

\section{Decay width of an impurity bounded in a quantum vortex\label{sec:Fermi-golden-rule}}

We motivate the introduction of the hydrodynamic densities of states
by obtaining the decay width between states of an impurity bounded
in a quantum vortex. 

We consider the BEC described by Hamiltonian~\eqref{eq:Form_Ham}
to be in the presence of atoms of a distinct species, described by
a Hamiltonian
\begin{equation}
\hat{H}_{\text{imp}}=\int\text{d}^{2}r\,\hat{\Psi}^{\dagger}(\mathbf{r})\left[-\frac{\hbar^{2}}{2M_{2}}\nabla^{2}+g_{12}\hat{\Phi}^{\dagger}(\mathbf{r})\hat{\Phi}(\mathbf{r})\right]\hat{\Psi}(\mathbf{r})\,,\label{eq:AppFermi_Ham-Impurity}
\end{equation}
that is, a field of dilute (i.e., non-interacting) atoms of mass $M_{2}$
interacting with BEC atoms by a contact potential of strength $g_{12}>0$;
we identify the atoms of mass $M_{2}$ as impurities with respect
to the BEC, since they are assumed to be dilute and of a distinct
species; the total Hamiltonian will be $\hat{H}=\hat{H}_{\text{BEC}}+\hat{H}_{\text{imp}}$.

Substituting the BEC fields for~\eqref{eq:Form_Bogol-presc} in~\eqref{eq:AppFermi_Ham-Impurity}
yields, to leading and sub-leading orders, 
\begin{align}
\hat{H}_{\text{imp}} & =\frac{\chi}{n_{0}\xi^{2}}\int\text{d}^{2}r\,\hat{\Psi}^{\dagger}h_{\text{imp}}\hat{\Psi}\nonumber \\
 & +\frac{\chi\gamma^{2}}{\left(n_{0}\xi^{2}\right)^{\frac{3}{2}}}\int\text{d}^{2}r\,\hat{\Psi}^{\dagger}\left(\overline{\Phi}_{\nu}\hat{\phi}+\Phi_{\nu}\hat{\phi}^{\dagger}\right)\hat{\Psi}+\text{\dots}\,,\label{eq:AppFermi_BEC-Impurity}
\end{align}
scaled to the natural units $n_{0}g$ and $\xi$, according to the
prescriptions in the main text, having introduced a Schrödinger operator
$h_{\text{imp}}=-\nabla^{2}+\gamma^{2}\left|\Phi_{\nu}\right|^{2}$
and parameters $\chi=M/M_{2}$ and $\gamma^{2}=M_{2}g_{12}/(Mg)$;
performing a further scaling $\hat{H}_{\text{imp}}\mapsto\frac{\chi}{n_{0}\xi^{2}}\hat{H}_{\text{imp}}$,
the total Hamiltonian can be written as 
\[
\hat{H}=F_{0}+\frac{1}{n_{0}\xi^{2}}\left(\hat{H}_{\text{B}}+\chi\hat{H}_{\text{imp}}\right)\,.
\]

Bosonic field operators $\hat{\phi}$ and $\hat{\phi}^{\dagger}$
are expanded as in Eq~\eqref{eq:Form_fluct-expand}, a basis of eigenfunctions
of $\mathcal{H}_{\text{B}}$, resulting
\[
\overline{\Phi}_{\nu}(\mathbf{r})\hat{\phi}(\mathbf{r})+\text{h.c.}=\sum_{m,n}\left(\zeta_{m,n}(\mathbf{r})\hat{b}_{m,n}+\overline{\zeta_{m,n}}(\mathbf{r})\hat{b}_{m,n}^{\dagger}\right)\,,
\]
\[
\zeta_{m,n}(\mathbf{r})=\overline{\Phi}_{\nu}(\mathbf{r})u_{m,n}(\mathbf{r})+\Phi_{\nu}(\mathbf{r})v_{m,n}(\mathbf{r})\,,
\]
where the identification $\lambda=(m,n)$ is made. Impurity field
operators $\hat{\Psi}$ and $\hat{\Psi}^{\dagger}$ can be expanded
in a basis of eigenfunctions of $h_{\text{imp}}$: the vortex profile
of the BEC density $\left|\Phi_{\nu}\right|^{2}$ \textendash{} along
with the condition that $g_{12}>0$ \textendash{} essentially guarantees
the existence of bound states of the impurities localized at the vortex
core~\cite{Braz:2020aa}; these are also eigenstates of angular
momentum in the plane, since the density $\left|\Phi_{\nu}\right|^{2}$
is cylindrically symmetric

For the purpose of this derivation, we consider a two-level truncated
basis comprised of the lowest-energy states of angular momenta $\ell\neq0$
and $\ell'=0$, i.e., we let $\hat{\Psi}(\mathbf{r})=\Psi_{0}(\mathbf{r})\hat{a}_{0}+\Psi_{\ell}(\mathbf{r})\hat{a}_{\ell}$,
yielding the effective Hamiltonian
\begin{align}
\hat{H}_{\text{eff}} & =\hat{H}_{\text{B}}+\Delta\left(\hat{a}_{\ell}^{\dagger}\hat{a}_{\ell}-\hat{a}_{0}^{\dagger}\hat{a}_{0}\right)\nonumber \\
 & +\sum_{m,n}\left(g_{\ell,0}^{(n)}\hat{b}_{m,n}+g_{0,\ell}^{(n)}\hat{b}_{m,n}^{\dagger}\right)\hat{a}_{\ell}^{\dagger}\hat{a}_{0}+\text{h.c.}\,,\label{eq:AppFermi_Ham-effect}
\end{align}
with $2\Delta$ the energy gap, where $g_{\ell,0}^{(n)}=\delta_{\ell,m}g_{\ell}^{(n)}$
and $g_{0,\ell}^{(n)}=\delta_{-\ell,m}g_{-\ell}^{(n)}$,
\[
g_{\ell}^{(n)}=\frac{\chi\gamma^{2}}{\sqrt{n_{0}\xi^{2}}}\int\text{d}^{2}r\,\overline{\Psi}_{\ell}\zeta_{\ell,n}\Psi_{0}\,;
\]
the selection rules above are made apparent from the fact that $L\zeta_{m,n}=m\zeta_{m,n}$
and $L\Psi_{\ell}=\ell\Psi_{\ell}$, with $L=-\text{i}\partial/\partial\varphi$
the two-dimensional angular momentum operator. An effective model
for the dynamics of the impurity coupled to the bosonic bath follows
from~\eqref{eq:AppFermi_Ham-effect} by projecting onto a single-particle
subspace of the impurity in the rotating-wave approximation and considering
only the bosonic modes of angular momentum $m=\ell$ :
\begin{align*}
\hat{H}_{\text{eff}} & =\sum_{n}\omega_{\ell,n}\hat{b}_{\ell,n}^{\dagger}\hat{b}_{\ell,n}+\Delta\sigma_{3}\\
 & +\sum_{n}\left(e^{-\text{i}t(\omega_{\ell,n}-\Delta)}g_{\ell}^{(n)}\hat{b}_{\ell,n}\sigma_{+}+\text{h.c.}\right)\,,
\end{align*}
with $\sigma_{+}$ the raising operator of impurity levels. A standard
approach to this problem is to employ the Wigner-Weisskopf approximation~\cite{Wang:1974aa},
which predicts a decay width
\[
\Gamma_{\ell\rightarrow0}=\pi\sum_{n}\bigl|g_{\ell}^{(n)}\bigr|^{2}\delta(\Delta-\omega_{\ell,n})\,.
\]
Then, considering that the impurity is localized in the vortex core
of a large BEC, the $\bigl|g_{\ell}^{(n)}\bigr|^{2}$ can be approximated
to rewrite the width as
\[
\Gamma_{\ell\rightarrow0}\approx\pi\frac{\chi^{2}\gamma^{4}}{n_{0}\xi^{2}}\int\text{d}^{2}r\,D_{\ell}^{(\text{H})}(\mathbf{r};\Delta)\bigl|\Psi_{\ell}(\mathbf{r})\bigr|^{2}\bigl|\Psi_{0}(\mathbf{r})\bigr|^{2}\,,
\]
where $D_{\ell}^{(\text{H})}(\mathbf{r};\Delta)=\sum_{n}\bigl|\zeta_{\ell,n}(\mathbf{r})\bigr|^{2}\delta(\Delta-\omega_{\ell,n})$
is a hydrodynamic AM-LDOS, Eq.~\eqref{eq:LDOS_hydro-ldos}.

\section{Computation of the radial BEC profile\label{sec:Computation-of-the}}

A box potential, such as considered in this work, endows the BEC wave
function with a nearly-homogeneous profile, with the exception of
an exponentially-fast depletion of the wave function at the border~\cite{pethick_smith_2008}.
On the one hand, as the size of the BEC increases, it becomes numerically
non-trivial to compute its full profile, seeing as the depletion becomes
steeper and more abrupt at the scale of the system, as illustrated
in Fig.~\ref{fig:fig-1}(a); on the other hand, provided that any
other non-homogeneous feature is sufficiently localized within the
bulk, that is, away from the border of the BEC, the computation of
its wave function and the computation of the wave function near the
border can be asymptotically separated. We employ this observation
to compute the solution of Eq.~\eqref{eq:Methods_GPE-reduced-vortex}
by introducing an ansatz of the form
\[
\phi_{\nu}(r)=\phi_{\Sigma}(r)\phi_{\partial\Sigma}(r)\,,
\]
where $\phi_{\Sigma}$ is the wave function in the bulk surface of
the quasi-2D BEC and $\phi_{\partial\Sigma}$ the wave function near
the border.

\begin{figure}
\centering{}\includegraphics{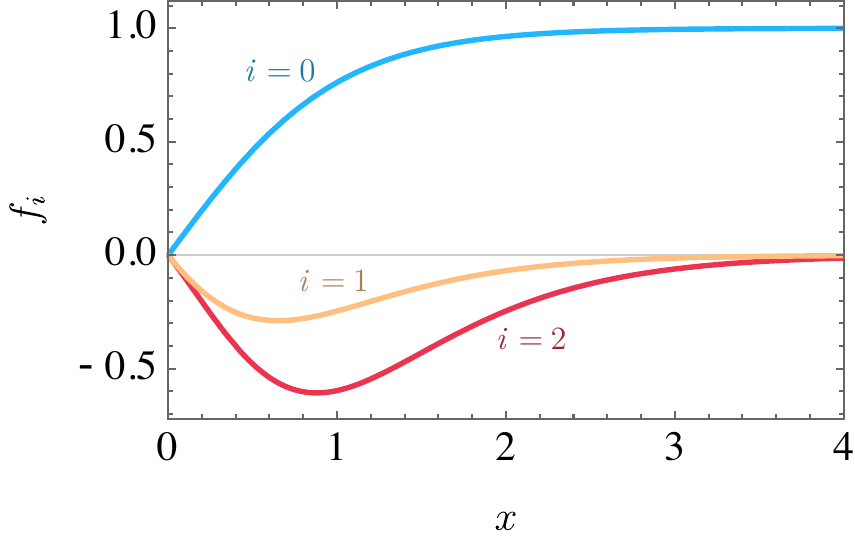}\caption{Asymptotic solutions $f_{i}$, for orders $i=0,1,2$, of Eq.~\eqref{eq:AppAsymp_f-eq}
for $\nu=1$ and $\mu\gtrsim1$.\label{fig:fig-9}}
\end{figure}

The bulk wave function $\phi_{\Sigma}$ is required to satisfy Eq.~\eqref{eq:Methods_GPE-reduced-vortex}
in absence of $V_{\text{tr}}$\textemdash that is, it is not required
to satisfy the boundary condition $\phi_{\nu}(R)=0$, but rather $\phi_{\Sigma}(r\rightarrow\infty)=\sqrt{\mu}$
by the asymptotic separation argument. The solution can be obtained
using imaginary-time evolution on a transformed radial coordinate
$\theta=2\arctan(r)$, with $\theta\in[0,\pi)$; we did not update
the chemical potential at each step of the evolution~\cite{Bao:2004aa}
but rather sampled the parametric $\mu$-dependence of the numerical
solutions. Further, because $V_{\text{p}}$ is exponentially localized
at the origin, the asymptotic expansion $\phi_{\Sigma}(r)=\sqrt{\mu}\left(1-\frac{\nu^{2}}{2\mu r^{2}}\right)+\text{\dots}$
holds for large $r$~\cite{Manton:2004aa}.

We now solve for $\phi_{\partial\Sigma}$ using an asymptotic approximation.
We begin by writing $\phi_{\partial\Sigma}(r)=f(x)$, for $i=0,1,2$,
with the coordinate $x=\sqrt{\mu/2}(R-r)$; the resulting equation
reads
\begin{align}
-\frac{1}{2}f''+\left(\frac{1}{\sqrt{2\mu}}\frac{1}{R}+\frac{x}{\mu R^{2}}\right)f' & +\nonumber \\
+\frac{\nu^{2}}{\mu R^{2}}f-\left[1-\left(1-\frac{\nu^{2}}{\mu R^{2}}\right)f^{2}\right]f & =\mathcal{O}(R^{-3})\,,\label{eq:AppAsymp_f-eq}
\end{align}
where we keep terms up to second order in $R^{-1}$; likewise, we
expand $f$ asymptotically to order $R^{-2}$:
\begin{equation}
f(x)=f_{0}(x)+R^{-1}f_{1}(x)+R^{-2}f_{2}(x)+\text{\dots}\,.\label{eq:AppAsymp_f-expand}
\end{equation}

Considering $x\in[0,\infty)$ by the asymptotic separation argument,
we require the boundary condition $f(0)=0$ and that $f(x)$ be bounded
as $x\rightarrow\infty$. To 0th order, the equation is $-\frac{1}{2}f_{0}''-\left(1-f_{0}^{2}\right)f_{0}=0$,
with solution $f_{0}(x)=\tanh(x)$, in agreement with Ref.~\cite{pethick_smith_2008}.
The 1st and 2nd-order equations are not worthwhile displaying here;
we mention only that we solved for the $f_{i}$ analytically with
the aid of symbolic computation software; results are plotted in Fig.~\ref{fig:fig-9}.

This method was validated against fully imaginary time-evolved solutions
for small system sizes ($R<70$). Because $R^{2}$ amounts to a dimensionless
coupling strength~\cite{Rokhsar:1997aa}, this approximation constitutes
a strong-coupling approximation. Moreover, further terms in the $R\gg1$
asymptotic expansion of $\phi_{\Sigma}$ can be obtained to compute
Eq.~\eqref{eq:AppAsymp_f-expand} to arbitrary order in $R^{-1}$.

\bibliographystyle{apsrev4-1}
\bibliography{References.bib}
\bibliographystyle{apsrev4-1}

\end{document}